# Spin-polarized light emitting diode using metal/insulator/semiconductor structures

T. Manago[a)] and H. Akinaga

*Research Consortium for Synthetic Nano-Function Materials Project (SYNAF),*
*National Institute of Advanced Industrial Science and Technology,*
*AIST Tsukuba Central 4, 1-1-1 Higashi, Tsukuba, Ibaraki 305-8562, Japan*

*Joint Research Center for Atom Technology (JRCAT),*
*AIST Tsukuba Central 4, 1-1-1 Higashi, Tsukuba, Ibaraki 305-0046, Japan*

We have succeeded in growing ferromagnetic metals (Co, Fe, and NiFe)/ $Al_2O_3$/ AlGaAs heterostructures with homogeneous and flat interfaces. The electro-luminescence (EL) from the light emitting diode (LED) consisting of the metal/insulator/semiconductor (MIS) structure depends on the magnetization direction of the ferromagnetic electrode at room temperature. This fact shows that a spin-injection from the ferromagnetic metal to the semiconductor is achieved. The spin-injection efficiency is estimated to be the order of 1 % at room temperature.

In recent years, a new field "spintronics", that is an idea to use the spin of electrons in electronic devices, progresses remarkably. Particularly a spin-injection from a ferromagnetic material into a semiconductor is one of the hot topics in this field. It was first realized using all-semiconducting heterostructured devices. Fiederling *et al.* injected spin-polarized electrons into a non-magnetic semiconductor GaAs through a II-VI diluted magnetic semiconductor $Be_xMn_yZn_{1-x-y}Se$, that was used as a spin aligner. They obtained a spin-injection efficiency of 90 % at 5 K.[1] Success of the spin-injection across the II-VI/III-V interface was also reported by Jonker *et al*.[2] Ohno *et al.* used a ferromagnetic semiconductor $Ga_{1-x}Mn_xAs$ as a spin-polarized hole source and injected the hole into GaAs with the efficiency of about 1% at 6 K.[3] Thus these experiments were succeeded at low temperature, since magnetic semiconductors does not work at room temperature. A lot of researchers make efforts to realize room-temperature operation of the spin-injection. Ferromagnetic metals (FM) are the likeliest candidates of the room-temperature spin-injecting sources. Theoretical prediction, however, showed that the limitation of the spin-injection efficiency from a metal into a semiconductor was less than 0.1 % due to a large conductance mismatch between them.[4] On the other hand, a tunneling process can inject spin-polarized electrons, since it is not affected by the conductance mismatch.[5] Zhu *et al.* reported the room-temperature spin injection using a Fe/GaAs Schottky barrier contact.[6] Since they used a reverse bias condition to inject electrons into GaAs, the spin-injection was done through the tunneling process across the Schottky barrier. The injection efficiency is about 2 %.

Metal-insulator-semiconductor (MIS) structures using GaAs have been investigated for a long time and various preparation methods were examined in the 1970's. The MIS structure has an advantage over the Schottky barrier in the spin-injection using the tunneling process. In this letter, preparation of a homogeneous $Al_2O_3$ tunnel barrier by a 2 steps deposition method on an AlGaAs/ GaAs/ AlGaAs light emitting diode (LED) is reported. We use ferromagnetic metals as a top electrode and demonstrate the spin-dependent electro-luminescence (EL) at room temperature.

The sample growth was performed using a multi-chamber molecular-beam epitaxy (MBE) with III-V growth, metal growth and oxidation chambers. The base pressure of these chambers was in the low $10^{-10}$ Torr range. The LED device structure was firstly grown on $p$-GaAs(001) substrate (Zn: ~ $1 \times 10^{18}$ cm$^{-3}$) by the III-V semiconductor MBE. The layer structure was $n$-$Al_{0.3}Ga_{0.7}As$ (Si: ~ $1 \times 10^{17}$ cm$^{-3}$) (50 nm)/ SI-$Al_{0.3}Ga_{0.7}As$ (50 nm)/ SI-GaAs (20 nm)/ SI-$Al_{0.3}Ga_{0.7}As$ (50 nm)/ $p$-$Al_{0.3}Ga_{0.7}As$ (Be: ~ $1 \times 10^{18}$ cm$^{-3}$) (200 nm)/ $p$-GaAs substrate. Subsequently, the $Al_2O_3$ tunnel barrier was made by the 2 steps deposition method, that is, 2ML Al deposition and 20 minutes $O_2$ oxidation, and reactive deposition of Al in an oxygen atmosphere. This process provides the best result in $Al_2O_3$ flatness and homogeneous. The thickness of the $Al_2O_3$ layer was 2 nm. Ferromagnetic electrode (Co, Fe and permalloy ($Ni_{80}Fe_{20}$)) with the thickness of 20 nm was grown on the $Al_2O_3$ layer at room temperature and the film was covered with an Au capping layer. Figure 1 shows a cross-sectional transmission electron microscopy (TEM) image of the LED structure (a) and a magnification of the MIS structure (b). Flat $Al_2O_3$ layer was realized. The homogeneity of our tunnel barriers on GaAs were checked by the thickness dependence of the junction resistivity.[7] The logarithm of the junction resistivity is proportional to the thickness $d$ of the $Al_2O_3$ tunnel barrier, which is expected from the WKB approximation. It indicates that the barrier is very homogeneous all over the sample. A schematic junction structure is shown in Fig. 1 (c). The junction area of 60 μmφ was fabricated using a conventional photolithography and an argon ion etching. A $SiO_2$ isolation layer was sputter deposited and a contact hole were made by liftoff. Metal pads with an opening window (48 μmφ) for an optical access to the junction were deposited on the contact hole. Back ohmic contacts were also formed by indium alloying at 200 °C.

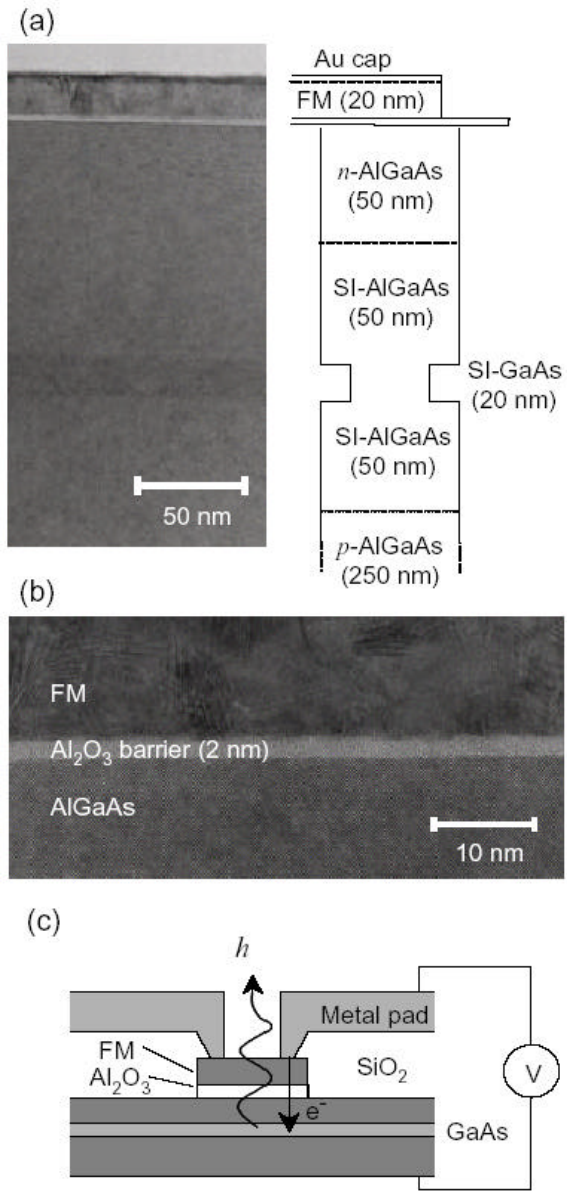

Fig. 1. (a) Cross-sectional TEM image of the LED structure. (b) Magnification of the MIS structure. Flat and homogeneous $Al_2O_3$ layer was realized. (c) The schematic junction structure. A junction area is 60 μmφ and an opening window is 48 μmφ for an optical access to the junction.

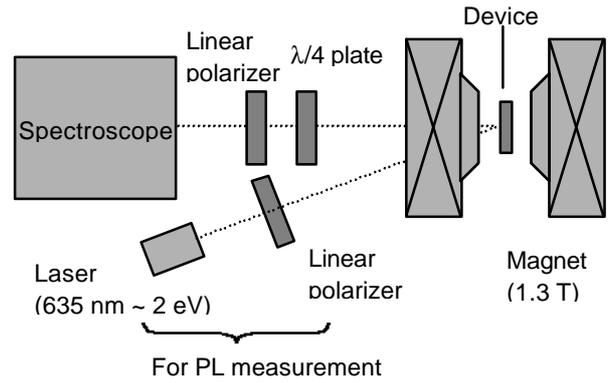

Fig. 2. Schematic view of the experimental setup for the detection of spin-dependent electro-luminescence. Magnetic field is applied up to 1.3 T. For PL measurement, a laser (λ = 635 nm) and a subsequent linear polarizer are used.

The spin-dependent EL measurement was performed by a spectroscope and a cooled charge-coupled device (CCD) (JASCO NRS-1000) with a λ/4 wave plate and an analyzer (linear polarizer) (Fig. 2). It can analyze left and right circular polarization of the emitted light from the devices. Both the applied magnetic field (up to 1.3 T) and the direction of emitted light are perpendicular to the sample surface. We used a DC measurement and detected the polarization difference by reversing of the applied field. The spin-polarization is defined as $P = (I_+ - I_-)/(I_+ + I_-)$, where $I_+$ and $I_-$ are intensities in the case of + and − of the applied magnetic fields (the wave plate was fixed). In order to estimate the magnetic circular dichroism (MCD) of the top ferromagnetic electrode, a photoluminescence (PL) measurement was also performed. The excitation photon energy was 2.0 eV (635 nm). The laser light passes through a linear polarizer and the top ferromagnetic electrode. Even if the incident light is slightly polarized by passing the ferromagnetic electrode, the light with the photon energy of 2 eV does not generate spin-polarized electrons.[8] Therefore we can ignore the MCD of the incident light and extract the MCD of the emitted light.

The EL spectrum of the LED with a Co electrode at room temperature is shown in Fig. 3 (a). The sample was biased at 2.8 V and the current is 22 mA. Devices begin to luminous at the bias of 1.8 V, however, at least the bias of more than 2.5 V is necessary to obtain good S/N ratio of the spin-dependent signal. For comparison, the spectrum of the device without the ferromagnetic electrode (Au electrode) is shown in fig. 3 (b). These spectra were observed around 1.42 eV with a shoulder. The shape of the spectrums reflects the energy level of the quantum well states in GaAs layer. Solid line and dotted line represent the spectrum with the applied magnetic field +1.3 T and −1.3 T, respectively. For the device with the Au electrode, it has no magnetic response. The spin-polarization $P$ is almost zero above 1.42 eV. The result of $P \approx 0$ is unchanged in the case of rotating the wave plate in 90-degree. Slight deviation at around the lower-energy of the peak edge (≤ 1.41 eV) is an artifact reflecting the peak shape originated from a lens of the apparatus. Therefore we regard the $P$ above 1.42 eV as an observed spin-polarization. For the device with the Co electrode, an intensity of the EL depends on the direction of the magnetic field. A 90-degree rotation of the wave plate completely exchanges the magnetic response of the EL intensity. The spin-polarization of the EL $P_{EL}$ (%) was estimated to be 3.5 %. Since the value includes the MCD of the top electrode, it is necessary that we extract effective spin-polarization by spin-injection. The PL measurement is the best tool of the purpose. Using an appropriate energy of the laser (2 eV), we can ignore the MCD of the incident light. The spin-polarization of the PL $P_{PL}$ (%) of the same sample film was also plotted in Fig. 3 (a). The value is estimated to be 2.7 %, which represents the MCD effect of the top electrode. Therefore the effective spin-polarization $P_{eff}$ is $P_{EL}$ (3.5 %) − $P_{PL}$ (2.7 %) = 0.8 %. Taking into account the four-fold degeneracy of the valence band, it indicates the spin-injection efficiency is $2P_{eff}$. Spin-polarization for other ferromagnetic electrodes was also investigated. For the device with a Fe electrode, the $P_{eff}$ is $P_{EL}$ (3.0 %) − $P_{PL}$ (2.5 %) = 0.5 %, and for the device with a NiFe electrode, the $P_{eff}$ is $P_{EL}$ (0.4 %) − $P_{PL}$ (0.2 %) = 0.2 %. The magnetic easy axis of the samples lies in plane. The ferromagnetic electrode did not saturate magnetically at 1.3 T because the magnetic field was applied normal to the surface. The spin-polarization increases in proportion to the magnetic field in the present study. The spin-injection efficiency is thought to reach about 2 % under the saturation magnetic field.

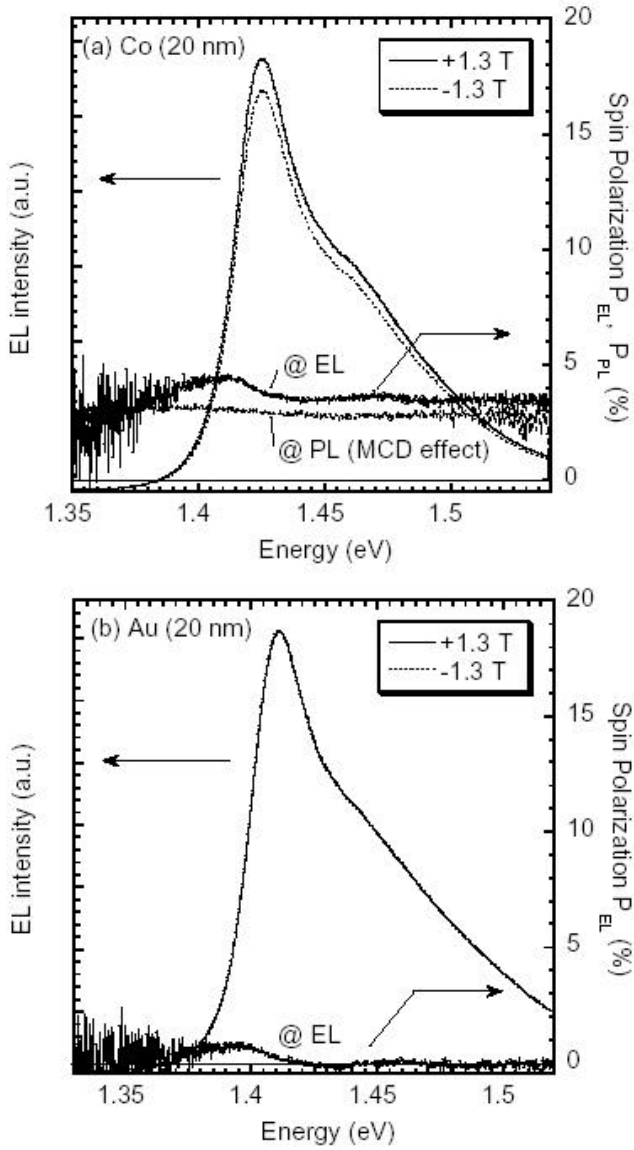

Fig. 3. (a) The EL intensity of the LED with the Co electrode (left axis). Solid and dotted lines represent the spectra with the applied magnetic fields of +1.3 T and −1.3 T. The spin polarization $P$ of EL and PL, $P_{EL}$ and $P_{EL}$, also plotted (right axis). The PL measurement represents the MCD effect of the top electrode. The difference between the $P_{EL}$ and the $P_{EL}$ corresponds to the spin-injection signal. All measurements were performed at room temperature. (b) The EL intensity of the LED with the Au electrode (left axis) and the spin polarization (right axis). It has no magnetic response. Slight deviation around the peak edge (≤ 1.41 eV) is an artifact come from a lens of the apparatus.

The spin-injection efficiency using the MIS structure is the same order of that of the Schottky barrier contact. In order to increase the spin-injection efficiency, reducing the bias across the tunnel barrier seems to be necessary. As is reported in a magnetic tunnel junction (MTJ), the tunnel magneto-resistance ratio (TMR) decreases with increasing sample bias.[9] Moreover, Moodera et al. reported that the junctions with NiFe electrodes showed stronger decrease in TMR than junctions with Co electrodes.[10] It may cause the smaller injection efficiency of our NiFe device than others since the spin-polarization of Co, Fe and NiFe is almost the same.[11] The decrease was considered to be due to the spin scattering in the tunnel barrier[12] and the magnon scattering in ferromagnetic electrode.[13] It was also reported that the bias dependence possibly originates from energy dependence of the spin-polarization around the Fermi level in the ferromagnetic electrodes.[14] There has been no tunnel barrier without degrade of the TMR by increasing bias so far. The bias dependence mechanism of the TMR is indispensable for higher spin-injection efficiency.

In conclusion, we prepared the LED with ferromagnetic metals/ $Al_2O_3$/ AlGaAs MIS structures. Flat and homogeneous tunnel barrier was realized on GaAs. Spin-injection from a ferromagnetic metal to a semiconductor was confirmed using MIS structure. Spin-dependent electro-luminescence of the devices showed the spin-polarization of the order of 1 % at room temperature. The injection efficiency was small, which is probably due to the high sample bias across the tunnel barrier.

Acknowledgements

This work, partly supported by the New Energy and Industrial Technology Development Organization (NEDO), was performed in JRCAT under a joint research agreement between Angstrom Technology Partnership (ATP) and National Institute for Advanced Industrial

Science and Technology (AIST). This work was also partly supported by NEDO under the Nanotechnology Materials Program.